\documentclass[12pt]{article}

\usepackage{latexsym}

\usepackage{graphicx}

\begin{document}

%\begin{flushright}
%Sofia University\\
%\end{flushright}
%%%%%%%%%%%%%%%%%%%%%%%%%%%%%%%%%%%%%%%%%%%%%%%%%%%%%%%%%%%%%%%%%%%

\title{Rotating dyonic dipole black rings:\\ exact solutions and thermodynamics}

\author{
     Stoytcho S. Yazadjiev \thanks{E-mail: yazad@phys.uni-sofia.bg}\\
{\footnotesize  Department of Theoretical Physics,
                Faculty of Physics, Sofia University,}\\
{\footnotesize  5 James Bourchier Boulevard, Sofia~1164, Bulgaria }\\
}

\date{}

\maketitle

\begin{abstract}
New rotating dyonic dipole black ring solutions are derived in
5D Einstein-dilaton gravity with antisymmetric forms. The black rings
are analyzed and their thermodynamics is discussed. New dyonic black string
solutions are also presented.
\end{abstract}

%%%%%%%%%%%%%%%%%%%%%%%%%%%%%%%%%%%%%%%%%%%%%%%%%%%%%%%%%%%%%%%%%%%

%\draft
\sloppy

\section{Introduction}

The growing interest in  higher dimensional gravity, and in higher dimensional black holes in particular,
reveals itself in different aspects. One of the directions of investigations is the construction
of exact  solutions both analytically and numerically (see for example  \cite{T}-\cite{REHR} and references therein.)

An interesting development in the black holes studies is the discovery of the black ring solutions
of the five-dimensional Einstein equations by Emparan and Reall \cite{ER1}, \cite{ER2}. These are
asymptotically flat solutions with an event horizon of topology $S^2\times S^1$ rather the
much more familiar $S^3$ topology. Moreover, it was shown in \cite{ER2} that both the black hole
and the the black ring can carry the same conserved charges, the mass and a single angular
momentum, and therefore there is no uniqueness theorem in five dimensions. Since the Emparan and
Reall's discovery many explicit examples of black ring solutions were found in various gravity
theories \cite{E}-\cite{P}. Elvang was able
to apply Hassan-Sen transformation to the solution \cite{ER2} to find a charged black ring in the
bosonic sector of the truncated heterotic string theory\cite{E}. This solution is the first
example of black rings with dipole charges depending, however, of the other physical parameters.
A supersymmetric black ring in
five-dimensional minimal supergravity was derived in \cite{EEMR1} and then generalized to the
case of concentric rings in \cite{GG1} and \cite{GG2}. A static black ring solution of the five
dimensional Einstein-Maxwell gravity was found by Ida and Uchida in \cite{IU}. In \cite{EMP}
Emparan derived "dipole black rings" in Einstein-Maxwell-dilaton (EMd) theory
in five dimensions. In this work Emparan showed that the black rings can exhibit novel
feature with respect to the black holes. The black rings can also carry independent nonconserved charges
which can be varied continuously without altering the conserved charges. This fact leads to
continuous non-uniqness. Following the same path that yields the
three-charge rotating black holes \cite{BLMPSV}-\cite{TSEY}, Elvang, Emparan and Figueras constructed a
seven-parameter family of supergravity solutions that describe non-supersymmetric black rings
and black tubes with three charges, three dipoles and two angular momenta \cite{EEF}.

The thermodynamics of the dipole black rings was studied first by Emparan in \cite{EMP} and by
Copsey  and Horowitz in \cite{CH}. Within the framework of the quasilocal
counterterm method, the thermodynamics of the dipole rings was discussed by Astefanesei and Radu \cite{AR}.
The first law of black rings thermodynamics in $n$-dimensional Einstein dilaton gravity with $(p+1)$-form field
strength was derived by Rogatko in \cite{ROG}. Static and asymptotically flat black ring solutions in five-dimensional
EMd gravity with arbitrary dilaton coupling parameter $\alpha$ were presented in \cite{KL}.
Asymptotically non-flat black rings immersed in external electromagnetic fields were found
and discussed  in \cite{O}, \cite{KL} and \cite{Y0}. Using solitonic technique, Mishima and Iguchi
derived the black ring solutions in 5D Einstein gravity \cite{MI},\cite{IM} (see also \cite{IM1}, \cite{TMY},\cite{TN}).
A systematical derivation of the asymptotically
flat static black ring solutions
in five-dimensional EMd gravity with an arbitrary dilaton coupling parameter was given in
\cite{Y}. In the same paper and in \cite{Y1}, the author systematically derived new type static and rotating
black ring solutions which are not asymptotically flat. Static dyonic black rings in 5D Einstein-dilaton gravity
with antisymmetric forms were found and studied in \cite{Y2}.

The aim of this paper is to present new rotating dipole black ring solutions in 5D Einstein-dilaton gravity
with antisymmetric forms and to study their thermodynamics. In order to achieved these goals we generalize
first the solution generating method of \cite{Y3} and \cite{Y4} in the presence of two antisymmetric forms.
Appling then  the solution  generating method we construct the exact rotating dipole black rings as well as
boosted black strings.

\section{Basic equations and solution generating}

We consider the action

\begin{eqnarray}\label{A}
S= {1\over 16\pi} \int d^{5}x \sqrt{-g}\left[R - 2g^{\mu\nu}\partial_{\mu}\varphi\partial_{\nu}\varphi
- {1\over 4} e^{-2\alpha\varphi} F_{\mu\nu}F^{\mu\nu} - {1\over 12}e^{-2\beta\varphi} H^{\mu\nu\lambda}H_{\mu\nu\lambda}  \right]
\end{eqnarray}

where $H = dB$ and $B$ is the Kalb--Ramond field . This action is the 5-dimensional version of
the action studied by Gibbons and Maeda \cite{GM}. Let us note that the 3-form field strength $H$ can be dualized\footnote{In the same way, the $2$-form $F$ can be dualized to a $3$-form ${\cal H}$ whose contribution  to the action is given by $-{1\over 12}  e^{2\beta\varphi} {\cal H}^{\mu\nu\lambda}{\cal H}_{\mu\nu\lambda}  $}  to 2-form field
strength ${\cal F}$ whose contribution to the action is given by $-{1\over 4}e^{2\beta\varphi}{\cal F}_{\mu\nu}{\cal F}^{\mu\nu}$. In other
words the theory we consider is equivalent to the Einstein-Maxwell-dilaton gravity with two distinct Maxwell fields and dilaton coupling
parameters. Particular examples of the action (\ref{A})(or its dual version)  with concrete values of the dilaton coupling
parameters arise from  string theory and supergravity via compactifications to five dimensions\footnote{As a result of the
compactifications, many additional fields come into play and one obtains rather complicated field models. In order
to obtain a simplified (truncated) action, as the one we consider here, we must suppress the additional fields by imposing certain selfconsistent
conditions (see for example \cite{SV}- \cite{KK}).             }.

The action (\ref{A}) yields the following field equations

\begin{eqnarray}\label{FE1}
R_{\mu\nu} &=& 2\partial_{\mu}\varphi \partial_{\nu}\varphi
+ {1\over 2}e^{-2\alpha\varphi} \left(F_{\mu\lambda}F_{\nu}^{\,\lambda} - {1\over 6} F_{\sigma\lambda}F^{\sigma\lambda} g_{\mu\nu}\right)
 \nonumber \\ &+& {1\over 4}e^{-2\beta\varphi} \left(H_{\mu\sigma\lambda}H_{\nu}^{\,\sigma\lambda}
- {2\over 9}H_{\rho\sigma\lambda}H^{\rho\sigma\lambda}g_{\mu\nu} \right),\nonumber  \\
\nabla_{\mu}\nabla^{\mu}\varphi &=& - {\alpha \over 8} e^{-2\alpha\varphi} F_{\sigma\lambda}F^{\sigma\lambda} -
{\beta\over 24 }e^{-2\beta\varphi}H_{\rho\sigma\lambda}H^{\rho\sigma\lambda}, \\
&&\nabla_{\mu}\left(e^{-2\alpha\varphi} F^{\mu\nu} \right) = 0 , \nonumber \\
&&\nabla_{\mu}\left(e^{-2\beta\varphi} H^{\mu\nu\lambda} \right) = 0 . \nonumber
\end{eqnarray}

In this paper we consider 5D spacetimes with  three commuting Killing vectors:
one timelike Killing vector $T$ and two spacelike Killing vectors $K_{1}$ and $K_{2}$.  We also assume
that the Killing vector $K_{2}$ is hypersurface orthogonal.

In adapted coordinates in which $K_{2}=\partial/\partial Y$, the spacetime
metric can be written in the form

\begin{equation}
ds^2 = e^{2u}dY^2 + e^{-u} h_{ij}dx^idx^j
\end{equation}

where $h_{ij}$ is a $4$-dimensional metric with Lorentz signature. Both $u$ and $h_{ij}$
depend on the coordinates $x^i$ only. The form field strengths are taken in the form\footnote{Throughout this paper
we denote the Killing vectors and their naturally corresponding 1-forms by the same letter. }

\begin{eqnarray}
F &=& 2d\Phi\wedge dY,\\
H &=& 2 e^{2\beta\varphi}\star (d\Psi \wedge dY ).
\end{eqnarray}

Performing dimensional reduction along the Killing vector $K_{2}$ we obtain the following effective 4D equations

\begin{eqnarray}
&&{\cal D}_{i}{\cal D}^{i} u = - {4\over 3} e^{-2\alpha\varphi - 2u}h^{ij}{\cal D}_{i}\Phi {\cal D}_{j}\Phi -
{4\over 3} e^{2\beta\varphi - 2u}h^{ij}{\cal D}_{i}\Psi {\cal D}_{j}\Psi ,\\
&&{\cal D}_{i}{\cal D}^{i} u = - \alpha  e^{-2\alpha\varphi - 2u}h^{ij}{\cal D}_{i}\Phi {\cal D}_{j}\Phi +
\beta e^{2\beta\varphi - 2u}h^{ij}{\cal D}_{i}\Psi {\cal D}_{j}\Psi ,\\
&&{\cal D}_{i}\left(e^{-2\alpha\varphi - 2u} {\cal D}^{i}\Phi \right) = 0,\\
&&{\cal D}_{i}\left(e^{2\beta\varphi - 2u} {\cal D}^{i}\Psi \right) = 0,\\
&&R(h)_{ij} = {3\over 2} {\cal D}_{i}u {\cal D}_{j}u + 2 {\cal D}_{i}\varphi {\cal D}_{j}\varphi +
2e^{-2\alpha\varphi - 2u} {\cal D}_{i}\Phi {\cal D}_{j}\Phi + 2e^{2\beta\varphi - 2u} {\cal D}_{i}\Phi {\cal D}_{j}\Psi,
\end{eqnarray}

where ${\cal D}_{i}$ and $R(h)_{ij}$ are the covariant derivative and the Ricci tensor with respect to the 4D metric $h_{ij}$.

These equations can be derived form the 4D action

\begin{eqnarray}\label{FDA}
S= {1\over 16\pi} \int d^4x \sqrt{-h} \left[R(h)  - {3\over 2 }h^{ij}{\cal D}_{i}u{\cal D}_{j}u
- 2 h^{ij}{\cal D}_{i}\varphi{\cal D}_{j}\varphi \nonumber \right. \\\left. - 2e^{-2\alpha\varphi -2u}h^{ij}{\cal D}_{i}\Phi{\cal D}_{j}\Phi
- 2e^{2\beta\varphi -2u}h^{ij}{\cal D}_{i}\Psi{\cal D}_{j}\Psi \right].
\end{eqnarray}

It turns out that the action (\ref{FDA}) possesses an important group of  symmetries when
the coupling  parameters $\alpha$ and $\beta$ satisfy\footnote{It has to be noted that this condition fixes values of $\alpha$
(and $\beta$) diferent from those predicted by string theory.}

\begin{equation}
\alpha_{*} \beta_{*} = 1
\end{equation}

where

\begin{equation}
\alpha_{*} = {\sqrt{3}\over 2} \alpha , \,\,\,  \beta_{*} = {\sqrt{3}\over 2} \beta.
\end{equation}

In order to find the symmetries of the action (\ref{FDA}),
we define the new fields $\xi= u + \alpha_{*}\varphi_{*}$ and $\eta = u - \beta_{*}\varphi_{*}$
and introduce the symmetric matrices

\begin{eqnarray}
M_{1} = e^{-\xi}\left(%
\begin{array}{cc}
  e^{2\xi} + (1 + \alpha^2_{*})\Phi^2_{*} &  \sqrt{1 + \alpha^2_{*}}\Phi_{*} \\
 \sqrt{1 + \alpha^2_{*}}\Phi_{*}  &  1 \\\end{array}%
\right) ,
\end{eqnarray}

\begin{eqnarray}
M_{2} = e^{-\eta}\left(%
\begin{array}{cc}
  e^{2\eta} + (1 + \beta^2_{*})\Psi^2_{*} &  \sqrt{1 + \beta^2_{*}}\Psi_{*} \\
 \sqrt{1 + \beta^2_{*}}\Psi_{*}  &  1 \\\end{array}%
\right) ,
\end{eqnarray}

where

\begin{eqnarray}
\varphi_{*} = {2\over \sqrt{3}} \varphi , \,\,\, \Phi_{*} = {2\over \sqrt{3}} \Phi, \,\,\, \Psi_{*} = {2\over \sqrt{3}} \Psi.
\end{eqnarray}

The matrixes $M_{1}$ amd $M_{2}$ satisfy  $\det M_{1}= \det M_{2}=1$.  Then the action (\ref{FDA})  can be written in the form

\begin{eqnarray}
S = {1\over 16\pi} \int d^4x\sqrt{-h} \left[R(h)
+ {3\over 4(1+ \alpha^2_{*})} h^{ij}Tr\left({\cal D}_{i}M_{1} {\cal D}_{j}M^{-1}_{1}\right) \nonumber \right. \\ \left.
+ {3\alpha^2_{*}\over 4(1+ \alpha^2_{*})} h^{ij}Tr\left({\cal D}_{i}M_{2} {\cal D}_{j}M^{-1}_{2}\right)\right].
\end{eqnarray}

It is now clear that the action  is invariant under the $SL(2,R)\times SL(2,R)$ group which acts as follows

\begin{eqnarray}
M_{1} \to AM_{1}A^{T} , \,\,\, M_{2} \to  BM_{2} B^{T}
\end{eqnarray}

where $A,B \in SL(2,R)$. Matrixes $M_{1}$
and $M_{2}$ parameterize the coset $SL(2,R)/SO(2)\times SL(2,R)/SO(2)$. Let us mention that similar non-linear $\sigma$-models
coupled to  the four-dimensional gravity and motivated by  Kaluza-Klein and extended supergravity theories 
were listed  in non-explicit form in \cite{BMG}.

Next step is to further reduce the effective $4D$ theory along the Killing vectors $T$ and $K_{1}$.
In this connection, it is useful to introduce the twist $\omega$ of the Killing vector $T$ defined by

\begin{eqnarray}\label{TD}
\omega = -{1\over 2} \star (h)\left(T\wedge dT \right)
\end{eqnarray}

were $\star(h)$ is the Hodge dual with respect to the metric $h_{ij}$.

One can show that the Ricci 1-form ${\Re}_{h}[T]$  defined by
\begin{equation}
{\Re}_{h}[T] = R_{ij}(h)T^{j}dx^{i} ,
\end{equation}

 satisfies

\begin{equation}
\star(h)\left( T\wedge {\Re}_{h}[T] \right) = d\omega .
\end{equation}

Obviously, in our case we have ${\Re}_{h}[T]=0$, i.e. $d\omega$=0.  Therefore there exists (locally) a  potential $f$ such that

\begin{equation}\label{EFORM}
\omega = df.
\end{equation}

In adapted coordinates for the Killing vectors $T=\partial/\partial t$ and $K_{1}=\partial/\partial X$,
and in the canonical coordinates $\rho$ and $z$ for the transverse space, the 4D metric $h_{ij}$ can be written in the form

\begin{eqnarray}
h_{ij}dx^idx^j = -e^{2U}\left(dt + {\cal A} dX \right)^2 + e^{-2U}\rho^2 dX^2 + e^{-2U}e^{2\Gamma}(d\rho^2 + dz^2).
\end{eqnarray}

For this form of the metric $h_{ij}$, combining (\ref{TD}) and (\ref{EFORM}), and  after some algebra we find
that  the twist potential $f$ satisfies

\begin{eqnarray}\label{TPS}
\partial_{\rho}f &=& -{1\over 2} {e^{4U}\over \rho} \partial_{z}{\cal A} ,\\
\partial_{z} f &=& {1\over 2} {e^{4U}\over \rho} \partial_{\rho}{\cal A}.
\end{eqnarray}

Before writing the 2D reduced equations  we shall introduce the symmetric matrix

\begin{eqnarray}
M_{3} = \left(%
\begin{array}{cc}
  e^{2U} + 4f^2e^{-2U} & 2fe^{-2U} \\
 2fe^{-2U} & e^{-2U} \\\end{array}%
\right)
\end{eqnarray}

with $\det M_{2}=1$. Then the 2D reduced EM equations read

\begin{eqnarray}
&&\partial_{\rho}\left[\rho \partial_{\rho }M^{-1}_{1} M_{1} \right] + \partial_{z}\left[\rho\partial_{z}M^{-1}_{1} M_{1} \right]
=0 , \\
&& \partial_{\rho}\left[\rho \partial_{\rho }M^{-1}_{2} M_{2} \right] + \partial_{z}\left[\rho\partial_{z}M^{-1}_{2} M_{2} \right]
=0 , \\
&& \partial_{\rho}\left[\rho \partial_{\rho }M^{-1}_{3} M_{3} \right] + \partial_{z}\left[\rho\partial_{z}M^{-1}_{3} M_{3} \right]
=0 , \\
&&\rho^{-1} \partial_{\rho} \Gamma = - {1\over 8} \left[Tr(\partial_{\rho}M_{3}\partial_{\rho}M^{-1}_{3})
- Tr(\partial_{z}M_{3}\partial_{z}M^{-1}_{3}) \right]   \nonumber \\
&& -  {3\over  8(1+ \alpha^2_{*})}  \left[Tr(\partial_{\rho}M_{1}\partial_{\rho}M^{-1}_{1})
- Tr(\partial_{z}M_{1}\partial_{z}M^{-1}_{1}) \right] \nonumber \\
&&-  {3\alpha^2_{*}\over  8(1+ \alpha^2_{*})}  \left[Tr(\partial_{\rho}M_{2}\partial_{\rho}M^{-1}_{2})
- Tr(\partial_{z}M_{2}\partial_{z}M^{-1}_{2}) \right] \\
&&\rho^{-1} \partial_{z}\Gamma = - {1\over 4} Tr(\partial_{\rho}M_{3}\partial_{z}M^{-1}_{3})  -
{3\over 4(1+ \alpha^2_{*})}  Tr(\partial_{\rho}M_{1}\partial_{z}M^{-1}_{1}) \nonumber \\
&&- {3\alpha^2_{*}\over 4(1+ \alpha^2_{*})}  Tr(\partial_{\rho}M_{2}\partial_{z}M^{-1}_{2}).
\end{eqnarray}

As a result we find that the "field variables" $M_{1}$, $M_{2}$  and $M_{3}$ satisfy the equations of three
$SL(2,R)/SO(2)$ $\sigma$-models in two dimensions, modified by the presence of the factor $\rho$.
The system equations for $\Gamma$
can be integrated, once a pair of solutions for the three $\sigma$-models are known. Therefore,
the problem of generating solutions to equations (\ref{FE1})  with the described symmetries reduces to
the solutions of the three $\sigma$-models.
It is well known  that the $\sigma$-model equations are completely integrable \cite{BZ1,BZ2}. Therefore
our theory with the imposed symmetries is complete integrable. The inverse scattering transform  method
can be used to generate solutions of the $\sigma$-model equations. However, in the present paper will proceed
in different way, namely we will follow the scheme of  \cite{Y3} and \cite{Y4} and will give a solution generating
method which allows us to construct new solutions from known solutions of the 5D vacuum Einstein equations.
Omitting the intermediate steps which are quite similar to those of \cite{Y3} and \cite{Y4}, we present the final result

{\bf Proposition.} {\it Let us consider three solutions of the vacuum
5D Einstein equations }

\begin{eqnarray}
ds_{E(i)}^2 = g^{E(i)}_{YY} dY^2 + g^{E(i)}_{00}\left(dt + {\cal A}^{(i)}_{E}dX \right)^{2} +
{\tilde g}^{E(i)}_{XX}dX^2 + g^{E(i)}_{\rho\rho} (d\rho^2 + dz^2)
\end{eqnarray}

{\it Then the following give a solution to the 5D equations  (\ref{FE1})   }

\begin{eqnarray}
ds^2 = \left[|g^{E(1)}_{00}| \sqrt{g^{E(1)}_{YY}}\right]^{2\over 1 + \alpha^2_{*}}
 \left[|g^{E(2)}_{00}| \sqrt{g^{E(2)}_{YY}}\right]^{2\alpha^2_{*}\over 1 + \alpha^2_{*}} dY^2  \nonumber \\
+  {\sqrt{g^{E(3)}_{YY}} \over \left[|g^{E(1)}_{00}| \sqrt{g^{E(1)}_{YY}}\right]^{1\over 1 + \alpha^2_{*}}
\left[|g^{E(2)}_{00}| \sqrt{g^{(E2)}_{YY}}\right]^{\alpha^2_{*}\over 1 + \alpha^2_{*}}}  \left[
g^{E(3)}_{00}\left(dt + {\cal A}^{(3)}_{E}dX \right)^2  + {\tilde g}^{E(3)}_{XX}dX^2  \nonumber \right. \\ \left.
 +  \left({|g^{E(1)}_{00}| g^{E(1)}_{YY}g^{E(1)}_{\rho\rho}\over
e^{2\Omega^{(1)}_{E} + {2\over 3}\Omega^{(3)}_{E} }} \right)^{3\over 1 + \alpha^2_{*}}  \left({|g^{E(2)}_{00}| g^{E(2)}_{YY}g^{E(2)}_{\rho\rho}\over
e^{2\Omega^{(2)}_{E} + {2\over 3}\Omega^{(3)}_{E} }} \right)^{3\alpha^2_{*}\over 1 + \alpha^2_{*}} g^{(3)}_{\rho\rho} (d\rho^2 + dz^2) \right],
\end{eqnarray}

\begin{eqnarray}
e^{2\alpha\varphi} = \left[ {|g^{E(1)}_{00}|^2 g^{E(1)}_{YY} \over |g^{E(2)}_{00}|^2 g^{E(2)}_{YY}}\right]^{\alpha^2_{*}\over 1 + \alpha^2_{*}} ,
\end{eqnarray}

\begin{eqnarray}
\Phi =  \pm {\sqrt{3}\over  \sqrt{1 + \alpha^2_{*}} } f^{(1)}_{E} + const ,
\end{eqnarray}

\begin{eqnarray}
\Psi = \pm {\sqrt{3}\alpha_{*}\over \sqrt{1+ \alpha^2_{*}} } f^{(2)}_{E} + const ,
\end{eqnarray}

{\it where  $f^{(i)}_{E}$ is a solution to the system  }

\begin{eqnarray}\label{TPS1}
\partial_{\rho}f^{(i)}_{E} &=& -{1\over 2} {(g^{E(i)}_{00})^2 g^{E(i)}_{YY}\over \rho} \partial_{z}{\cal A}^{(i)}_{E} ,\\
\partial_{z} f^{(i)}_{E} &=& {1\over 2} {(g^{E(i)}_{00})^2 g^{E(i)}_{YY}\over \rho} \partial_{\rho}{\cal A}^{(i)}_{E},
\end{eqnarray}

{\it and $\Omega^{(i)}_{E}$ satisfy }

\begin{eqnarray}\label{OS1}
\rho^{-1}\partial_{\rho}\Omega^{(i)}_{E} &=& {3\over 16}\left[\left(\partial_{\rho} \ln\left( g^{E(i)}_{YY}\right)\right)^2
- \left(\partial_{z} \ln \left(g^{E(i)}_{YY}\right)\right)^2 \right],\\
\rho^{-1}\partial_{z}\Omega^{(i)}_{E} &=&
{3\over 8} \partial_{\rho} \ln \left(g^{E(i)}_{YY}\right)\partial_{z}\ln\left( g^{E(i)}_{YY}\right).
\end{eqnarray}

\section{Exact solutions }

In order to generate dipole black rings in the framework of the theory under consideration, we shall follow
the scheme outlined in \cite{Y3} and \cite{Y4}.

We take three copies of the neutral black ring solution with different parameters:
the first solution is with parameters $\{\lambda_{1},\nu,{\cal R}\}$, the second  with parameters
$\{\lambda_{2},\nu,{\cal R}\}$, while the third is parameterized by
$\{\lambda_{3},\nu,{\cal R}\}$:

\begin{eqnarray}
ds^2_{E(i)} = -{F_{\lambda_{i}}(y)\over F_{\lambda_{i}}(x)} \left(dt + C(\nu,\lambda_{i})
{\cal R}{1+y\over F_{\lambda_{i}}(y)}d\psi \right)^2 \nonumber\\
+ {{\cal R}^2\over (x-y)^2 }F_{\lambda_{i}}(x)\left[-{G(y)\over F_{\lambda_{i}}(y)}d\psi^2 - {dy^2\over G(y)} + {dx^2\over G(x)}
+ {G(x)\over F_{\lambda_{i}}(x)}d\phi^2 \right]
\end{eqnarray}

where

\begin{equation}
F_{\lambda_{i}}(x) = 1 + \lambda_{i} x ,\,\,\, G(x) = (1-x^2)(1+\nu x),
\end{equation}

and

\begin{equation}
C(\nu,\lambda_{i})= \sqrt{\lambda_{i}(\lambda_{i}-\nu){1 +\lambda_{i} \over 1-\lambda_{i}}}.
\end{equation}

The coordinates $x$ and $y$ vary in the range

\begin{equation}
 -1\le x \le 1 , \,\,\,\, -\infty < y \le -1 .
\end{equation}

It should be also noted that in the case under consideration
the Killing vectors are denoted by

\begin{equation}
K_{1} = {\partial/\partial \psi} , \,\,\,\, K_{2} = {\partial/\partial \phi}.
\end{equation}

The neutral black ring solution has already been written  in canonical coordinates in \cite{HAR},
that is why we present here the final formulas:

\begin{eqnarray}
|g^{E(i)}_{00}| &=& {(1+\lambda_{i})(1-\nu)R_{1} + (1-\lambda_{i})(1+\nu)R_{2}
-2(\lambda_{i} - \nu)R_{3}
- \lambda_{i}(1-\nu^2){\cal R}^2  \over (1+\lambda_{i})(1-\nu)R_{1}
+ (1-\lambda_{i})(1+\nu)R_{2}
-2(\lambda_{i} - \nu)R_{3}
+ \lambda_{i}(1-\nu^2){\cal R}^2 } , \nonumber \\
g^{E(i)}_{\Phi\Phi} &=& {(R_{3}+z - {1\over 2}{\cal R}^2 )(R_{2} - z
+ {1\over 2}{\cal R}^2\nu)
\over R_{1} - z - {1\over 2}{\cal R}^2\nu }  \nonumber \\ &=& {(R_{1} + R_{2} + \nu{\cal R}^2)  (R_{1} - R_{3}
+ {1\over 2}(1+ \nu){\cal R}^2) (R_{2} + R_{3} - {1\over 2}(1 - \nu){\cal R}^2)
\over {\cal R}^2 ((1-\nu)R_{1} - (1+\nu)R_{2} -2\nu R_{3}) } ,\nonumber \\
 g^{E(i)}_{\rho\rho} &=&  [(1+\lambda_{i})(1-\nu)R_{1} + (1-\lambda_{i})(1+\nu)R_{2}
 -2(\lambda_{i} - \nu)R_{3}
+ \lambda_{i}(1-\nu^2){\cal R}^2 ] \nonumber \\
&& \times {(1-\nu)R_{1} + (1+\nu)R_{2} + 2\nu R_{3}
\over  8(1-\nu^2)^2 R_{1}R_{2}R_{3}} , \\
{\cal A}^{(i)}_{E} &=& {-2 C(\nu,\lambda_{i}) {\cal R} (1-\nu)
[R_{3} -R_{1} + {1\over 2}{\cal R}^2 (1+\nu)] \over
(1+\lambda_{i})(1-\nu)R_{1} + (1-\lambda_{i})(1+\nu)R_{2} -2(\lambda_{i} - \nu)R_{3}
- \lambda_{i}(1-\nu^2){\cal R}^2 },  \nonumber
\end{eqnarray}

where

\begin{eqnarray}
R_{1} =\sqrt{\rho^2 + (z + {\nu\over 2}{\cal R}^2)^2 } , \\
R_{2} =\sqrt{\rho^2 + (z - {\nu\over 2}{\cal R}^2)^2 }, \\
R_{3} = \sqrt{\rho^2 + (z - {1\over 2}{\cal R}^2)^2 }.
\end{eqnarray}

The functions $\Omega^{(i)}_{E}$ and $f^{(i)}_{E}$ were found in \cite{Y3} and \cite{Y4} and they are given by

\begin{eqnarray}
e^{{8\over 3} \Omega_{E}^{(i)}} &=& { [(1-\nu)R_{1} + (1+\nu)R_{2}
+ 2\nu R_{3}]^2\over 8(1-\nu^2)^2R_{1}R_{2}R_{3} }
g^{E(i)}_{\phi\phi} ,\\
f^{(i)}_{E} &=& {(1-\nu) {\cal R} C(\nu,\lambda_{i}) [R_{1} - R_{3} +
{1\over 2}(1+\nu ) {\cal R}^2 ] \over (1+\lambda_{i})(1-\nu)R_{1} +
(1-\lambda_{i})(1+\nu)R_{2} + 2(\nu-\lambda_{i})R_{3}
+ \lambda_{i}(1-\nu^2){\cal R}^2 } \nonumber .
\end{eqnarray}

Since the metric function $g^{E}_{\phi\phi}$ does not depend on the parameter $\lambda$ we have
 $g^{E(1)}_{\phi\phi}=g^{E(2)}_{\phi\phi}=g^{E(2)}_{\phi\phi}$ which implies $\Omega^{(1)}_{E}= \Omega^{(2)}_{E}=\Omega^{(2)}_{E}$.
Taking this into account we find the following dyonic solution

\begin{eqnarray}
ds^2 = |g^{E(1)}_{00}|^{2\over 1 + \alpha^2_{*}} |g^{E(2)}_{00}|^{2\alpha^2_{*}\over 1 + \alpha^2_{*}} g^{E(3)}_{YY} dY^2
+ |g^{E(1)}_{00}|^{-{1\over 1 + \alpha^2_{*}}} |g^{E(2)}_{00}|^{-{\alpha^2_{*}\over 1 + \alpha^2_{*}}}
\left[ g^{E(3)}_{00}\left(dt + {\cal A}^{(3)}_{E} d\phi \right)^2  \right.  \nonumber \\ \left.+
\left({g^{E(1)}_{00}g^{E(1)}_{YY} g^{E(1)}_{\rho\rho}\over e^{{8\over 3 }\Omega^{(1)}_{E}}} \right)^{3\over 1+ \alpha^2_{*}}
 \left({g^{E(2)}_{00}g^{E(2)}_{YY} g^{E(2)}_{\rho\rho}\over e^{{8\over 3 }\Omega^{(2)}_{E}}} \right)^{3\over 1+ \alpha^2_{*}}
g^{E(3)}_{\rho\rho} (d\rho^2 + dz^2) \right] ,
\end{eqnarray}

\begin{eqnarray}
e^{2\alpha\varphi} = \left[ {|g^{E(1)}_{00}|\over |g^{E(2)}_{00}|}\right]^{2\alpha^2_{*}\over 1+ \alpha^2_{*}}
\end{eqnarray}

\begin{eqnarray}
\Phi =  \pm {\sqrt{3}\over  \sqrt{1 + \alpha^2_{*}} } f^{(1)}_{E} + const ,
\end{eqnarray}

\begin{eqnarray}
\Psi = \pm {\sqrt{3}\alpha_{*}\over \sqrt{1+ \alpha^2_{*}} } f^{(2)}_{E} + const ,
\end{eqnarray}

It is more convenient to
present the solution in coordinates in which it takes  simpler form. Such coordinates are the so-called
$C$-metric coordinates given by

\begin{eqnarray}
\rho = {{\cal R}^2 \sqrt{-G(x)G(y)}\over (x-y)^2 } ,\,\,\,
z = {1\over 2} {{\cal R}^2(1-xy)(2+\nu x + \nu y )\over (x-y)^2 }.
\end{eqnarray}

Performing this coordinate change we find

\begin{eqnarray}
ds^2 &=& - {F_{\lambda_{3}}(y)\over F_{\lambda_{3}}(x)} \left({F_{\lambda_{1}}(y)\over F_{\lambda_{1}}(x)}\right)^{-{1\over 1+ \alpha^2_{*}}}
\left({F_{\lambda_{2}}(y)\over F_{\lambda_{2}}(x)}\right)^{-{\alpha^2_{*}\over 1+ \alpha^2_{*}}}
\left( dt + C(\nu,\lambda_{3})
{\cal R}{1+y\over F_{\lambda_{3}}(y)}d\psi \right)^2 \nonumber \\ &&+
 \left({F_{\lambda_{1}}(y)\over F_{\lambda_{1}}(x)}\right)^{-{1\over 1+ \alpha^2_{*}}}
\left({F_{\lambda_{2}}(y)\over F_{\lambda_{2}}(x)}\right)^{-{\alpha^2_{*}\over 1+ \alpha^2_{*}}}
\left[ - {{\cal R}^2 \over (x-y)^2} {F_{\lambda_{3}}(x) G(y)\over F_{\lambda_{3}}(y) } d\psi^2  \right. \nonumber \\ && \left. +
\left( F_{\lambda_{1}}(y) \right)^{3\over 1+ \alpha^2_{*}} \left( F_{\lambda_{2}}(y) \right)^{3\alpha^2_{*}\over 1+ \alpha^2_{*}}
{{\cal R}^2 F_{\lambda_{3}}(x)\over (x-y)^2 } \left({dx^2 \over G(x)} - {dy^2\over G(y)} \right)\right]  \nonumber \\
&&+  \left({F_{\lambda_{1}}(y)\over F_{\lambda_{1}}(x)}\right)^{2\over 1+ \alpha^2_{*}}
\left({F_{\lambda_{2}}(y)\over F_{\lambda_{2}}(x)}\right)^{2\alpha^2_{*}\over 1+ \alpha^2_{*}} {{\cal R}^2 G(x)\over (x-y)^2 } d\phi^2 ,\\
e^{2\alpha\varphi} &=& \left[{F_{\lambda_{1}}(y)F_{\lambda_{2}}(x) \over F_{\lambda_{1}}(x)F_{\lambda_{2}}(y)} \right]^{2\alpha^2_{*}\over 1+ \alpha^2_{*} }, \\
\Phi &=& \pm {\sqrt{3}C(\nu,\lambda_{1})\over 2\sqrt{1+ \alpha^2_{*}} } {{\cal R} (1+ x)\over F_{\lambda_{1}}(x) } + const ,\\
\Psi &=& \pm {\sqrt{3}\alpha_{*} C(\nu,\lambda_{2})\over 2\sqrt{1+ \alpha^2_{*}} } {{\cal R} (1+ x)\over F_{\lambda_{2}}(x) } + const .
 \end{eqnarray}

It is useful to give the explicit expressions of the $2$-forms $B$ and ${\cal B}$
where $H=dB$ and\footnote{We recall that  ${\cal H}$ is the $3$-form dual to $F$.} ${\cal H}= d{\cal B}$:

\begin{eqnarray}
B_{t\psi} &=& \pm {\sqrt{3}\alpha_{*}C(\nu,\lambda_{2})\over \sqrt{1+ \alpha^2_{*}} } {\cal R} {1+y\over F_{\lambda_{2}}(y) } + const ,\\
{\cal B}_{t\psi} &=&  \pm {\sqrt{3}C(\nu,\lambda_{1})\over \sqrt{1+ \alpha^2_{*}} } {\cal R} {1+y\over F_{\lambda_{1}}(y) } + const .
\end{eqnarray}

For $\lambda_{2}=0$ we recover the dipole solutions in EMd gravity.

As we will see below the found solutions are specified by their mass $M$, their angular momentum $J_{{\tilde \psi}}$
and two dipole charges ${\cal Q}_{1}$ and ${\cal Q}_{2}$ which are not conserved charges. The dipole charges are independent of the mass and the angular momentum and are classically continuous parameters. Therefore our solutions 
exhibit 2-fold continuous non-uniqueness.

\section{Analysis of the solutions}

It is not difficult to see that $F_{\lambda_{i}}(x)=0$ and $F_{\lambda_{i}}(y)=0$ (i=1,2) correspond to curvature singularities.
In order to get rid of them we impose

\begin{equation}
\lambda_{1} = - \mu_{1} , \,\,\, \lambda_{2} = -\mu_{2}
\end{equation}

where

\begin{equation}
0 \le \mu_{1} < 1 , \,\,\, 0\le \mu_{2} < 1.
\end{equation}

The parameters $\lambda_{3}$ and $\nu$ satisfy

\begin{equation}
0<\nu \le \lambda_{3}<1.
\end{equation}

Further, the analysis of the solutions is similar to that for the neutral black ring. The possible conical singularities at $x=-1$ and $y=-1$ are
avoided by setting

\begin{eqnarray}\label{CSC1}
\Delta \psi = \Delta \phi = 2\pi {\sqrt{1-\lambda_{3}}   \over 1- \nu } (1+\mu_{1})^{3\over 2(1+ \alpha^2_{*})} (1 +\mu_{2})^{3\alpha^2_{*}\over 2(1+ \alpha^2_{*})}
\end{eqnarray}

for the periods of the coordinates $\psi$ and $\phi$.

The balance between the forces in the ring is achieved when no conical singularity is present at $x=1$. This requires that

\begin{equation}\label{CSC2}
\Delta \phi = 2\pi {\sqrt{1 + \lambda_{3}}\over 1 + \nu} (1- \mu_{1})^{3\over 2(1+ \alpha^2_{*})} (1-  \mu_{2})^{3\alpha^2_{*}\over 1+ \alpha^2_{*}} .
\end{equation}

The conditions (\ref{CSC1}) and (\ref{CSC2}) are simultaneously satisfied only if

\begin{equation}
\left({1-\nu \over 1+ \nu }\right)^2 = {1-\lambda_{3}\over 1 + \lambda_{3} } \left({1 + \mu_{1} \over 1- \mu_{1} } \right)^{3 \over 1 + \alpha^2_{*} } \left({1 + \mu_{2} \over 1- \mu_{2} } \right)^{3\alpha^2_{*} \over 1 + \alpha^2_{*} } .
\end{equation}

The solutions have a regular horizon of topology $S^{2}\times S^{1}$ at $y=-1/\nu$ and ergosurface with the same topology at $y=-1/\lambda_{3}$. Also, there is a curvature singularity at $y=-\infty$.

We can see that the metric  is asymptotically flat using the the change of the coordinates. Let us introduce the new coordinates $r$, $\theta$, ${\tilde \psi}$ and ${\tilde \phi}$ given by

\begin{eqnarray}
r\cos\theta &=& {\cal R}_{*} {\sqrt{y^2 -1}\over x-y },\\
r\sin\theta &=& {\cal R}_{*} {\sqrt{1-x^2}\over x-y },\\
{\tilde \psi } &=& {2\pi \over \Delta \psi}\psi ,\\
{\tilde \phi } &=& {2\pi \over \Delta \phi}\phi ,
\end{eqnarray}

where

\begin{equation}
{\cal R}_{*} = (1+ \mu_{1})^{3\over 2(1+ \alpha^2_{*})} (1+ \mu_{2})^{3\alpha^2_{*}\over 2(1+ \alpha^2_{*})}
{\sqrt{1-\lambda_{3}}\over \sqrt{1 -\nu} } {\cal R}.
 \end{equation}

We note that the new angle coordinates ${\tilde \psi}$ and ${\tilde \phi}$ have the canonical periodicity $2\pi$.
In terms of the new coordinates and for $r\to \infty$ (i.e. $x=y=-1$) we obtain

\begin{equation}
ds^2 \approx - dt^2 + dr^2 + r^2\cos^2\theta d{\tilde \psi}^2 + r^2\sin^2\theta d{\tilde \phi}^2 .
\end{equation}

\section{Thermodynamics of the dyonic dipole black rings }

Profound discussion of the thermodynamics of the dipole black rings in Einstein-dilaton gravity with one antisymmetric form was given in \cite{CH} ( see also  \cite{ROG}). The generalization in the case of Einstein-dilaton gravity with two (or more )
antisymmetric forms is straightforward. That is why we refer reader to \cite{CH} and \cite{ROG} for details.

To study the thermodynamics we have to find the conserved charges  of our system first.
The mass of the solutions is found from the asymptotic of $g_{00}$. In our case we have

\begin{equation}
g_{00}\approx - \left[1 - \left({2\lambda_{3}\over 1 -\lambda_{3}}  + {1\over 1+ \alpha^2_{*}} {2\mu_{1} \over 1 + \mu_{1}} + {\alpha^2_{*}\over 1+ \alpha^2_{*}} {2\mu_{2} \over 1 + \mu_{2}}  \right){{\cal R}^2_{*}\over r^2 } \right]
\end{equation}

whence we determine the mass

\begin{eqnarray}
M &=& {3\pi {\cal R}^2_{*}\over 4 } \left({\lambda_{3}\over 1 -\lambda_{3}}  + {1\over 1+ \alpha^2_{*}} {\mu_{1} \over 1 + \mu_{1}} + {\alpha^2_{*}\over 1+ \alpha^2_{*}} {\mu_{2} \over 1 + \mu_{2}}  \right) , \\
&=& {3\pi {\cal R}^2\over 4 } {(1+ \mu_{1})^{3\over 1+ \alpha^2_{*} } (1+ \mu_{2})^{3\alpha^2_{*}\over 1+ \alpha^2_{*} } \over  1- \nu } \left(\lambda_{3} + {1\over 1+ \alpha^2_{*}} {\mu_{1}(1-\lambda_{3}) \over 1 + \mu_{1}} + {\alpha^2_{*}\over 1+ \alpha^2_{*}} {\mu_{2}(1-\lambda_{3}) \over 1 + \mu_{2}}  \right) \nonumber .
\end{eqnarray}

In the same way, from the asymptotic

\begin{eqnarray}
g_{t{\tilde \psi}} \approx {2C(\nu,\lambda_{3}) \over (1-\lambda_{3})\sqrt{1-\nu} } {{\cal R}^{3}_{*}\cos^2\theta\over r^2 }
\end{eqnarray}

we find for the angular momentum

\begin{eqnarray}
J_{{\tilde \psi}} &=& - {\pi {\cal R}^3_{*}\over 2 } {C(\nu,\lambda_{3}) \over (1-\nu)\sqrt{1-\lambda_{3}} } \\
&=& - {\pi {\cal R}^3\over 2 } {C(\nu,\lambda_{3}) \sqrt{1-\lambda_{3}}\over (1-\nu)^2} (1+ \mu_{1})^{9\over 2(1+ \alpha^2_{*}) } (1+ \mu_{2})^{9\alpha^2_{*}\over 2(1+ \alpha^2_{*}) } \nonumber .
\end{eqnarray}

The angular velocity of the horizon can be easily calculated and the result is

\begin{eqnarray}
\omega_{h} = - {1\over {\cal R} } {(\lambda_{3} -\nu) \over \sqrt{1-\lambda_{3}} C(\nu,\lambda_{3}) }
(1+ \mu_{1})^{-{3\over 2(1+ \alpha^2_{*})}} (1+ \mu_{2})^{-{3\alpha^2_{*}\over 2(1+ \alpha^2_{*})}}.
\end{eqnarray}

The area of the horizon time-slice is found by a straightforward calculation

\begin{eqnarray}
{\cal A}_{h} = 8\pi^2 {\cal R}^3 (1+\mu_{1})^{3\over 1 + \alpha^2_{*}} (1+\mu_{2})^{3\alpha^2_{*}\over 1 + \alpha^2_{*}}
{\sqrt{\lambda_{3} (1- \lambda^2_{3}) }\over (1+ \nu)(1-\nu)^2 } (\mu_{1} + \nu)^{3\over 2(1+ \alpha^2_{*})}
(\mu_{2} + \nu)^{3\alpha^2_{*}\over 2(1+ \alpha^2_{*})} .
\end{eqnarray}

The entropy is one quarter of the horizon area

\begin{equation}
S = {1\over 4} {\cal A}_{h} .
\end{equation}

The temperature can be obtained via the surface gravity

\begin{eqnarray}
T = {1\over  4\pi {\cal R}} \sqrt{1-\lambda_{3} \over \lambda_{3}(1+\lambda_{3}) } \nu(1+\nu) (\mu_{1} + \nu)^{-{3\over 2(1+ \alpha^2_{*})}}
(\mu_{2} + \nu)^{-{3\alpha^2_{*}\over 2(1+ \alpha^2_{*})}}.
\end{eqnarray}

The dipole charges are defined by

\begin{eqnarray}
{\cal Q}_{1} &=& {1\over 4\pi} \oint_{S^2} e^{-2\beta\varphi}\star H , \\
{\cal Q}_{2} &=& {1\over 4\pi} \oint_{S^2} e^{2\alpha\varphi} \star {\cal H} ,
\end{eqnarray}

where the integral is over any  $S^2$ which can be continuously deformed to an $S^2$ on the horizon.
These dipole charges are found to be

\begin{eqnarray}
{\cal Q}_{1} &=& {\pm} {\sqrt{3}\alpha_{*}C(\nu,-\mu_{2})\over \sqrt{1 + \alpha^2_{*}} } {{\cal R} \over 1 -\mu_{2}}
{\sqrt{1-\lambda_{3}}\over 1- \nu } (1+ \mu_{1})^{3\over 2(1+ \alpha^2_{*})} (1+ \mu_{2})^{3\alpha^2_{*}\over 2(1+ \alpha^2_{*})} ,\\
{\cal Q}_{2} &=& {\pm} {\sqrt{3}C(\nu,-\mu_{1})\over \sqrt{1 + \alpha^2_{*}} } {{\cal R} \over 1 -\mu_{1}}
{\sqrt{1-\lambda_{3}}\over 1- \nu } (1+ \mu_{1})^{3\over 2(1+ \alpha^2_{*})} (1+ \mu_{2})^{3\alpha^2_{*}\over 2(1+ \alpha^2_{*})} .
\end{eqnarray}

Further we  define the potentials $\chi_{1}$ and $\chi_{2}$ as the difference between the values of ${\cal B}$  and $B$ at infinity and on the horizon

\begin{eqnarray}
\chi_{1} &=& {\pi\over 2} \left[{\cal B}_{t{\tilde \psi}}(x=y-1) - {\cal B}_{t{\tilde \psi}}(y=-{1\over \nu})  \right], \\
\chi_{2} &=& {\pi\over 2} \left[B_{t{\tilde \psi}}(x=y-1) -  B_{t{\tilde \psi}}(y=-{1\over \nu})  \right],
\end{eqnarray}

which give

\begin{eqnarray}
\chi _{1} &=& \pm {\pi\sqrt{3} \alpha_{*}C(\nu, -\mu_{2} )\over 2\sqrt{1+ \alpha^2_{*}} } {\cal R} {\sqrt{1-\lambda_{3}}\over \mu_{2} + \nu } (1+ \mu_{1})^{3\over 2(1+ \alpha^2_{*})} (1+ \mu_{2})^{3\alpha^2_{*}\over 2(1+ \alpha^2_{*})},\\
\chi _{2} &=& \pm {\pi\sqrt{3} \alpha_{*}C(\nu, -\mu_{1} )\over 2\sqrt{1+ \alpha^2_{*}} } {\cal R} {\sqrt{1-\lambda_{3}}\over \mu_{1} + \nu } (1+ \mu_{1})^{3\over 2(1+ \alpha^2_{*})} (1+ \mu_{2})^{3\alpha^2_{*}\over 2(1+ \alpha^2_{*})} .
\end{eqnarray}

A straightforward calculation show that the dyonic dipole black rings satisfy a Smarr-like relation

\begin{eqnarray}
M = {3 \over  2} \left(TS + J_{{\tilde \psi}}\omega_{h}\right) + {1\over 2}\chi_{1} {\cal Q}_{1} + {1\over 2}\chi_{2}{\cal Q}_{2}.
\end{eqnarray}

From the results of \cite{CH} and \cite{ROG} generalized to our case
in which two antisymmetric forms are present, follows that  the firs law

\begin{eqnarray}
dM = TdS + \omega_{h} dJ_{{\tilde \psi}} + \chi_{1}d{\cal Q}_{1}  + \chi_{2}d{\cal Q}_{2}
\end{eqnarray}

is satisfied. Of course, the first law can be also checked by direct calculations.

\section{Boosted dyonic black strings }

Here we consider the case when the radius of the ring grows very large. In this case, in the limit ${\cal R}\to \infty$, as one expects, we obtain boosted straight dyonic black strings. In  order to see that let us define

\begin{eqnarray}
r_{0}= \nu {\cal R}, \,\,\,\,  \cosh^2(\sigma) = {\lambda_{3} \over \nu} , \,\,\, \mu_{i} {\cal R}= r_{0} \sinh^2(\gamma_{i})
\end{eqnarray}

and

\begin{eqnarray}
r = - {{\cal R}\over y},\,\,\,  \cos\theta = x, \,\,\, \eta = {\cal R}\psi .
\end{eqnarray}

Taking the limit ${\cal R}\to \infty$ , $\lambda, \nu, \mu_{i} \to 0$ and keeping $r_{0},\sigma,\gamma_{i}$ and $r,\theta, \eta$ finite we obtain the following solution

\begin{eqnarray}
ds^2 &=&  h^{- {1\over 1+ \alpha^2_{*} }}_{1} h^{- {\alpha^2_{*}\over 1+ \alpha^2_{*} }}_{2}
\left[ - {\hat f}\left(dt - {r_{0} \cosh(\sigma)\sinh(\sigma)\over r {\hat f} }d\eta \right)^2
+ {f\over {\hat f}  }  d\eta^2 \right] \nonumber \\
&& +  h^{{2\over 1+ \alpha^2_{*} }}_{1} h^{{2\alpha^2_{*}\over 1+ \alpha^2_{*} }}_{2} \left({dr^2\over f} + r^2d\Omega^2_{2}  \right) , \\
e^{2\alpha\phi} &=& \left({h_{1}\over h_{2}} \right)^{2\alpha^2_{*}\over 1 + \alpha^2_{*}} ,\\
\Phi &=& \pm {\sqrt{3} \over \sqrt{1 + \alpha^2_{*}}} r_{0} \cosh(\gamma_{1})\sinh(\gamma_{1}) (1+ \cos\theta),\\
\Psi &=& \pm {\sqrt{3} \alpha_{*}\over \sqrt{1 + \alpha^2_{*}}} r_{0} \cosh(\gamma_{2})\sinh(\gamma_{2}) (1+ \cos\theta),
\end{eqnarray}

where

\begin{eqnarray}
f &=& 1 - {r_{0}\over r}, \\
{\hat f} &=&  1 - {r_{0}\cosh^2(\sigma) \over r},
\end{eqnarray}

\begin{eqnarray}\label{HFUNK}
h_{i} &=& 1 + {r_{0}\sinh^2(\gamma_{i})\over r } .
\end{eqnarray}

The physical quantities characterizing the dyonic black string solutions can be found as  limits of the
corresponding quantities for the black rings.

The black string solutions can be derived via the solution generating method presented above. It is instructive
to give this derivation explicitly.
Our starting point is the vacuum solution

\begin{eqnarray}\label{VBBSS}
ds^2_{E}= - {\hat f} \left(dt - {r_{0} \cosh(\sigma)\sinh(\sigma)\over r {\hat f} }d\eta \right)^2  +  {f\over {\hat f} }  d\eta^2 +  {dr^2\over f} + r^2d\Omega^2_{2}.
\end{eqnarray}

This solution can be written in the canonical coordinates $\rho$ and $z$ by the coordinate change

\begin{eqnarray}\label{CCC}
r= L + {r_{0}\over 2 } ,\,\,\, z= L \cos\theta \\
\cos\theta = {2\Delta \over r_{0}} , \,\,\, \rho^2 = \left(L^2 - {r^2_{0}\over 4 } \right) \sin^2\theta
\end{eqnarray}

where

\begin{eqnarray}
L = {1\over 2} \left[\sqrt{\rho^2 + (z+ {r_{0}\over 2 })^2} + \sqrt{\rho^2 + (z- {r_{0}\over 2 })^2} \right],\\
\Delta = {1\over 2} \left[\sqrt{\rho^2 + (z+ {r_{0}\over 2 })^2} - \sqrt{\rho^2 + (z- {r_{0}\over 2 })^2} \right].
\end{eqnarray}

Then we obtain

\begin{eqnarray}
g^{E}_{00} &=& - {L +  {r_{0}\over 2 }   - r_{0}\cosh^2(\sigma)  \over L + {r_{0}\over 2 }},\\
{\tilde g}^{E}_{\eta\eta} &=& {L + {r_{0}\over 2 } \over L +  {r_{0}\over 2 }   - r_{0}\cosh^2(\sigma) },\\
{\cal A}_{E} &=& - {r_{0}\sinh(\sigma)\cosh(\sigma)\over L +  {r_{0}\over 2 }   - r_{0}\cosh^2(\sigma)  }, \\
g^{E}_{\rho\rho} &=& {(L + {r_{0}\over 2 })^2\over L^2 - \Delta^2 } .
\end{eqnarray}

For $u_{E}$ we have

\begin{eqnarray}
u_{E} = \ln(\rho) + {1\over 2}\ln \left[{L+ {r_{0}\over 2 }\over L - {r_{0}\over 2}} \right].
\end{eqnarray}

After  tedious calculations we find

\begin{eqnarray}
e^{{8\over 3 }\Omega_{E}} = {(L + {r_{0}\over 2 })^3  \over  (L^2 - \Delta^2) (L- {r_{0}\over 2})} \rho^2 .
\end{eqnarray}

Further we consider three copies of the vacuum solution (\ref{VBBSS}): the first with parameters
$\{r_{0}^{(1)}=-r_{0}, \sigma^{(1)}= \gamma_{1} \}$, the second with $\{r_{0}^{(2)}=-r_{0}, \sigma^{(2)}= \gamma_{2} \}$ and
the third parameterized by $\{r_{0}^{(3)}=r_{0}, \sigma^{(3)}= \sigma \}$. According to the proposition the metric
and the dilaton field are given by

\begin{eqnarray}
ds^2 &=& h^{{2\over 1 +\alpha^2_{*}}}_{1} h^{{2\alpha^2_{*}\over 1 +\alpha^2_{*}}}_{2} g^{E(3)}_{\phi\phi} d\phi^2  +
h^{-{1\over 1 +\alpha^2_{*}}}_{1} h^{-{\alpha^2_{*}\over 1 +\alpha^2_{*}}}_{2} \left[g^{E(3)}_{00} \left(dt + {\cal A}^{(3)}_{E}d\eta \right)^2 + {\tilde g}^{E(3)}_{\eta\eta}d\eta^2  \right. \nonumber \\
&& + \left.
\left({|g^{E(1)}_{00}|g^{E(1)}_{\phi\phi} g^{E(1)}_{\rho\rho} \over e^{2\Omega^{(1)}_{E}  + {2\over 3} \Omega^{(3)}_{E}} } \right)^{3\over 1 + \alpha^2_{*}} \left({|g^{E(2)}_{00}|g^{E(2)}_{\phi\phi} g^{E(2)}_{\rho\rho} \over e^{2\Omega^{(2)}_{E}  + {2\over 3} \Omega^{(3)}_{E}} } \right)^{3\alpha^2_{*}\over 1 + \alpha^2_{*}}
g^{E(3)}_{\rho\rho}(d\rho^2 + dz^2) \right] ,  \\
e^{2\alpha\varphi} &=& \left({h_{1}\over h_{2} } \right)^{2\alpha^2_{*}\over 1 + \alpha^2_{*}} ,
\end{eqnarray}

where

\begin{equation}
h_{i} = |g^{E(i)}_{00}|\sqrt{{g^{E(i)}_{\phi\phi}\over g^{E(3)}_{\phi\phi} }} ,\,\,\, i= 1,2  \, .
\end{equation}

In explicit form we have

\begin{eqnarray}
h_{i} &=& {L - {r_{0}\over 2 } + r_{0}\cosh^2(\gamma_{i})\over L + {r_{0}\over 2 } } ,\\
\left({|g^{E(i)}_{00}|g^{E(i)}_{\phi\phi} g^{E(1)}_{\rho\rho} \over e^{2\Omega^{(i)}_{E}  + {2\over 3} \Omega^{(3)}_{E}} } \right) &=& {L - {r_{0}\over 2 } + r_{0}\cosh^2(\gamma_{i})\over L + {r_{0}\over 2 } } = h_{i} ,
\end{eqnarray}

where we have taken into account that $L^{(1)}=L^{(2)}=L^{(3)}=L$ and $\Delta^{(1)}=\Delta^{(2)}= -\Delta^{(3)}=- \Delta$.
In this way we find for the metric

\begin{eqnarray}
ds^2 = h^{-{1\over 1 +\alpha^2_{*}}}_{1} h^{-{\alpha^2_{*}\over 1 +\alpha^2_{*}}}_{2} \left[g^{E(3)}_{00} \left(dt + {\cal A}^{(3)}_{E}d\eta \right)^2 + {\tilde g}^{E(3)}_{\eta\eta}d\eta^2  \right] \nonumber \\
+ h^{{2\over 1 +\alpha^2_{*}}}_{1} h^{{2\alpha^2_{*}\over 1 +\alpha^2_{*}}}_{2}
\left[ g^{E(3)}_{\rho\rho}(d\rho^2 + dz^2) + g^{E(3)}_{\phi\phi} d\phi^2 \right]
\end{eqnarray}

Performing the coordinate change $(\rho,z)\to (r,\theta)$ using (\ref{CCC})
we obtain

\begin{equation}
h_{i} = 1 + {r_{0}\sinh^2(\gamma_{i})\over r}
\end{equation}

which  coincides with (\ref{HFUNK}).  Obviously, in $r,\theta$ coordinates the metric obtained via the
solution generating method coincides with that found by the limiting procedure.

It remains to find $f^{(i)}_{E}$ from

\begin{eqnarray}
\partial_{\rho}f^{(i)}_{E} &=&  {1\over 2}\rho  {r_{0}\sinh(\gamma_{i})\cosh(\gamma_{i})\over L^2 - {r^2_{o}\over 4 } } \partial_{z}L ,\\
\partial_{z}f^{(i)}_{E} &=&  - {1\over 2}\rho  {r_{0}\sinh(\gamma_{i})\cosh(\gamma_{i})\over L^2 - {r^2_{o}\over 4 } } \partial_{\rho}L .
\end{eqnarray}

The solution of this system  is

\begin{eqnarray}
f^{(i)}_{E} =  - r_{0}\sinh(\gamma_{i})\cosh(\gamma_{i}) \left(1+ {z\over L} \right) + const .
\end{eqnarray}

Hence, we find

\begin{eqnarray}
\Phi &=& \pm {\sqrt{3}\over \sqrt{1 + \alpha^2_{*}} }f^{(1)}_{E} + const = \pm r_{0}\sinh(\gamma_{1})\cosh(\gamma_{1}) \left(1+ \cos\theta \right)
+ const , \\
\Psi &=&  \pm {\sqrt{3}\alpha_{*}\over \sqrt{1 + \alpha^2_{*}} }f^{(2)}_{E} + const = \pm r_{0}\sinh(\gamma_{2})\cosh(\gamma_{2})
\left(1+ \cos\theta \right) +  const ,
\end{eqnarray}

which coincide with those obtained  by the limiting procedure.

\section{Conclusion}

In this paper we generalized the solution generating method of \cite{Y3} and \cite{Y4} in the case when two
antisymmetric forms are present in the equations of the dilaton gravity.  New solutions describing dyonic
dipole black rings were constructed and their thermodynamics was discussed. New dyonic black string  solutions were also
presented.

Let us finish with some words about future investigations. It would be interesting  to  find black ring solutions with
independent  net and dipole charges in Einstein-Maxwell-dilaton gravity
(and more generally  in Einstein-dilaton gravity with antisymmetric forms ).
The construction of such solutions, however, needs  modification of the solution generating method.
The results will be presented in future publications
\footnote{Some particular results have been already  presented on  seminars at the Institute
of Nuclear  Research and Nuclear Energy, Bulgarian Academy of Sciences, 18 and 25 May, 2006.   }.

\section*{Acknowledgements}

I would like to thank Prof. Ivan Todorov for the invitation
to give seminars at the Institute of Nuclear Research and Nuclear Energy, Bulgarian Academy of Sciences, where a part
of this work was presented.  My thanks also go to I. Stefanov  for reading the manuscript.
This work was partially supported by the Bulgarian National Science Fund under Grant MUF04/05 (MU 408)
and the Sofia University Research Fund under Grant No.60.

\end{document}